# Low Power Oriented CMOS Circuit Optimization Protocol

A. Verle, X. Michel, N. Azemard, P. Maurine, D. Auvergne
LIRMM, UMR CNRS/Université de Montpellier II,
(C5506),161 rue Ada, 34392 Montpellier, France
azemard, pmaurine, auvergne@lirmm.fr

## Abstract

*Low power oriented circuit optimization consists in selecting the best alternative between gate sizing, buffer insertion and logic structure transformation, for satisfying a delay constraint at minimum area cost.*

*In this paper we used a closed form model of delay in CMOS structures to define metrics for a deterministic selection of the optimization alternative. The target is delay constraint satisfaction with minimum area cost. We validate the design space exploration method, defining maximum and minimum delay bounds on logical paths. Then we adapt this method to a "constant sensitivity method" allowing to size a circuit at minimum area under a delay constraint. An optimisation protocol is finally defined to manage the trade-off performance constraint – circuit structure. These methods are implemented in an optimization tool (POPS) and validated by comparing on a 0.25μm process, the optimization efficiency obtained on various benchmarks (ISCAS'85) to that resulting from an industrial tool.*

## 1. Introduction

Current trade-off between speed, power and area is the obliged way for designing fast and reasonably powered modern integrated circuits. This can be achieved with circuit simulators and critical path analysis tools to modify iteratively the size of the transistors until the complete constraint satisfaction [1-4]. More general speed-up techniques involve buffer insertion [5-6] and logic transformation [7]. Although efficient for speeding-up combinational paths, these techniques may have different impacts in the resulting power dissipation or area. Gate sizing is area (power) expensive and, due to the resulting capacitive loading effects, may slow down adjacent upward paths. This implies iterative timing verifications. Buffer insertion preserves path interaction but is only efficient for relatively highly loaded nodes. Path structure modification implies a characterization of the gate efficiency and must deterministically be selected as a preprocessing step. To compare these alternatives it is necessary to evaluate the performance of the different implementations. Without using any robust indicator, selecting between all these different techniques for the various gates of a library is NP complex and induces more iterative attempts, which are processing time explosive.

A deterministic selection of speed-up technique must be based on the characterization of the feasible speed on a path, on the determination of the critical nodes and the characterization of the gate efficiency to the load and of its sensitivity to the sizing or buffering alternatives.

The main contribution of this paper is to define different metrics for gate and path characterization, to be used as efficient indicators for the selection of one of the best optimization alternative.

Section 2 provides background material for this study. The optimization alternative with structure conservation is presented and validated in section 3. The proposed optimization method with structure modification is detailed and validated in section 4, in which the resulting optimization protocol is presented, before to conclude in section 5.

## 2. Background

Using current path optimization tools [8] requires large CPU times and too significant computer resources to manage the complexity of actual circuits [9]. The uncertainty in routing capacitance estimation imposes to use many iterations or to consider very large safety margin resulting in oversized designs.

### 2.1. Optimization tool

As a solution to these drawbacks, we have developed an analysis and performance optimization tool based on an accurate representation of the physical abstraction of the layout (POPS: Performance Optimization by Path Selection) [10]. It offers possibilities in analyzing and optimizing combinatorial circuit paths in submicronic technologies.

This tool allows to consider an user specified limited number of paths [11-12], for easy application and validation of the different path optimization criteria. The delay model implemented in this tool is based on a design oriented closed-form representation of the timing performance, allowing to obtain for any logic gate, in its environment, an accurate evaluation of its switching delay and output transition time.

## 2.2. Delay Model

Real delay computation must consider finite input transition and I/O coupling [13]. We capture the effect of the input-to-output coupling and the input slope effect in the delay as

$$t_{HL}(i) = \frac{v_{TN}}{2}\tau_{INLH}(i-1) + (1+\frac{2C_M}{C_M+C_L})\frac{\tau_{outHL}}{2}(i) \quad (1)$$
$$t_{LH}(i) = \frac{v_{TP}}{2}\tau_{INHL}(i-1) + (1+\frac{2C_M}{C_M+C_L})\frac{\tau_{outLH}}{2}(i)$$

where $v_{TN,P}$ represents the reduced value ($V_T/V_{DD}$) of the threshold voltage of the N,P transistors. $\tau_{INHL,LH}$, $\tau_{outHL,LH}$ are the input and output transition time duration, respectively. $C_M$ is the coupling capacitance between the input and output nodes, that can be evaluated as one half the input capacitance of the P(N) transistor for input rising (falling) edge, respectively or directly calibrated from SPICE simulation.

The general expression of the transition time has been developed in [14] as

$$\tau_{outHL} = \tau \cdot S_{HL} \cdot \frac{C_L}{C_{IN}} \quad (2)$$
$$\tau_{outLH} = \tau \cdot S_{LH} \cdot \frac{C_L}{C_{IN}}$$
$$S_{HL} = (1+k) \cdot DW_{HL} \quad (3)$$
$$S_{LH} = R \cdot \frac{(1+k)}{k} \cdot DW_{LH}$$

where $\tau$ is a time unit that characterizes the process. $C_L$ and $C_{IN}$ represent, respectively, the output load and the gate input capacitance. $S_{HL,LH}$ represents the symmetry factor of the falling, rising edges. R represents, for an identical load and drive capacitance, the ratio of the current value available in N and P transistors, k is the P/N configuration ratio and $D_{WHL,LH}$ the gate logical weight defined by the ratio of the current available in an inverter to that of a serial array of transistors [14].

If eq.2,3 are quite similar to the logical effort expressions [4], they only represent the transition time expression. The delay is given by (1) that completely captures the input-to-output coupling and the input transition time effect on the delay. Using these expressions to define metrics for optimization, we always consider that the resulting implementation is in the fast input control range [14].

As shown from eq. (1-3) the delay on a bounded combinatorial path is a convex function and these expressions can easily be used to determine the best condition for path optimization under delay constraint.

By bounded combinatorial path we signify that the path input gate capacitance is fixed by the load constraint imposed on the latch supplying the path. This implies that the path terminal load is completely determined by the total input capacitance of the gates or registers controlled by this path. This guarantees the convexity of the delay on this path.

## 3. Optimization with structure conservation

The goal of gate sizing is to determine the optimum size for path delay constraint satisfaction at the minimum area/power cost. For that an essential parameter to be considered is the feasibility of the constraint imposed on the path. The target of this section is twofold: defining the delay bounds of a given path and determining a way for distributing a delay constraint on this path at the minimum area/power cost.

### 3.1. Constraint feasibility

Without indication on the feasibility of a constraint any iterative method may infinitely loop with no chance to reach a solution. For that, in order to verify the feasibility of a constraint, we explore the path optimization space by defining the max and min delay bounds (Tmax, Tmin) of this path. It is clear that if the delay constraint value is lower than the minimum delay achievable on this path, whatever is the optimization procedure, there is no way to satisfy the constraint without path modification. These bounds are of great importance in first defining the optimization alternative.

Theoretically and without gate size limitation, no upper delay bound can be defined on a path. To obtain a pseudo-upper bound (at minimum area) we just consider a realistic configuration in which all the gates are implemented with the minimum available drive.

The definition of the lower bound has been the subject of numerous proposals. For ideal inverters without parasitic loading, the minimum is reached when all the inverters have an equal tapering factor that can be easily calculated from a first order delay representation [7,15]. Applying the explicit representation given in (1) to a bounded combinatorial path, the inferior delay bound is easily obtained by canceling the derivative of the path delay with respect to the input capacitance of the gates. This results in a set of link equations

$$C_{IN}^2(i) = \frac{A_i}{A_{i-1}} \cdot \left(C_{IN}(i+1) + C_{par}(i)\right) \cdot C_{IN}(i-1) \quad (4)$$

where (i) specifies the rank of the gate, $C_{par}(i)$ is the gate (i) output parasitic capacitance and the $A_i$ correspond to the design parameters involved in (1,2).

As shown, the size of gate (i) depends on that of (i+1) and (i-1). This is exactly what we are looking for. Instead to solve the corresponding set of equations we prefer to use an iterative approach starting from a local solution defined with $C_{IN}(i-1)$ equal to the minimum available drive ($C_{REF}$). Then processing backward from the output, where the terminal load is known, to the input, we can easily determine an initial solution. Applying this solution in (4) we can reach, after few iterations, the minimum of delay achievable on the path. An illustration of the evolution of these iterations is given in Fig.1.

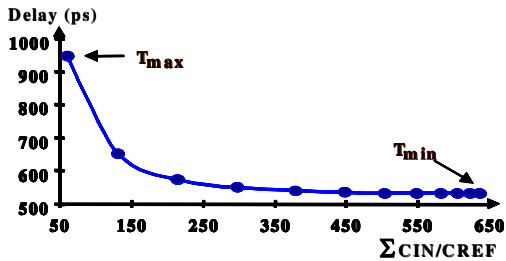

**Fig.1. Illustration of the sensitivity of the path delay to the gate sizing.**

We can easily verify that the final value, Tmin is conserved whatever is the initial solution, ie the $C_{REF}$ value. This method has been implemented in POPS. Validation has been obtained by comparing on the longest path of different ISCAS'85 benchmarks (process CMOS, 0.25μm) the minimum delay value, obtained from the proposed method, to that reached by an industrial tool (AMPS from Synopsis). The delay values are obtained from SPICE simulations of the corresponding path implementations. Fig.2 illustrates the resulting comparison that demonstrates the accuracy of the proposed method, for each case the minimum value obtained is lower than that resulting from a pseudo-random sizing technique.

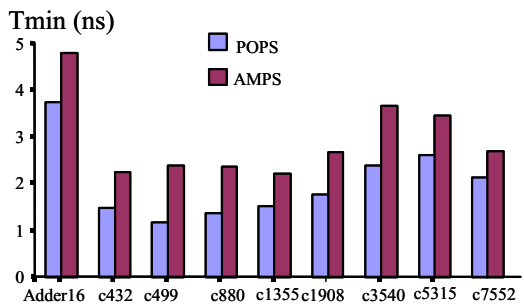

**Fig.2. Comparison of the minimum delay value (Tmin) determined with POPS and AMPS.**

The determination of each path delay bounds gives possibilities in verifying the feasibility of the constraint.

For a delay constraint value higher than the minimum bound, the optimization alternative to be chosen is transistor sizing with structure conservation. Otherwise a structure modification of the path must be considered. Next step is to develop a fast technique allowing to efficiently distribute the constraint on the path.

### 3.2 Constraint distribution.

Various methods can be used. The simplest method is the Sutherland method [4], directly deduced from the Mead's optimization rule of an ideal inverter array [15]: the same delay constraint is imposed on each element of the path. If this supplies a very fast method for distributing the constraint, this is at the cost of an over-sizing of the gates with an important logical weight value

We propose a new method based on the gate sensitivity to the sizing, that can be directly deduced from (4), and is illustrated in Fig.3.

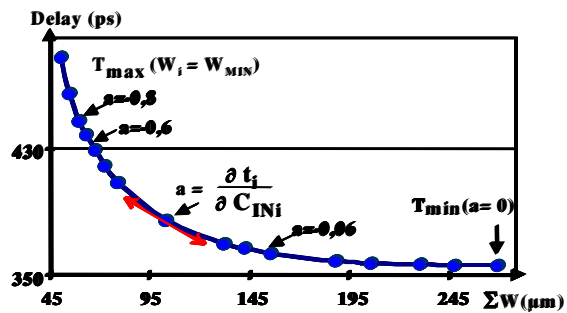

**Fig.3. Application of the constant sensitivity method to an 11 gate path.**

This Figure represents the variation of the path delay to the gate sizing. Each point has been obtained by imposing the same value of each partial derivative:

$$\frac{\partial T}{\partial C_{IN}(i)} = a \qquad (5)$$

"a" = 0 corresponds to the previously determined minimum. Varying the value of this coefficient from 0 to a large negative value allows the exploration of the full design space. This is obtained by solving:

$$A_{i-1} \cdot \frac{1}{C_{i-1}} - A_i \cdot \frac{C_{i+1} + C_{Pi}}{C_i^2} = a \qquad (6)$$
$$A_i \cdot \frac{1}{C_i} - A_{i+1} \cdot \frac{C_{i+2} + C_{Pi+1}}{C_{i+1}^2} = a.......$$

The solution of (6) supplies the sizes to be imposed to the input gate capacitance for satisfying the value of the sensitivity coefficient "a". Note that a path delay is associated to each value of the parameter "a". Few iterations on the "a" value allows a quick satisfaction of the delay constraint.

This method has been implemented in POPS and validated on different ISCAS circuits. In Fig.4 we compare the final area, given as the sum of the transistor widths (ΣW), necessary to implement the critical path of each circuit under an identical hard constraint (Tc = 1.2Tmin), using POPS and AMPS. As shown the equal sensitivity method results in a smaller area/ power implementation.

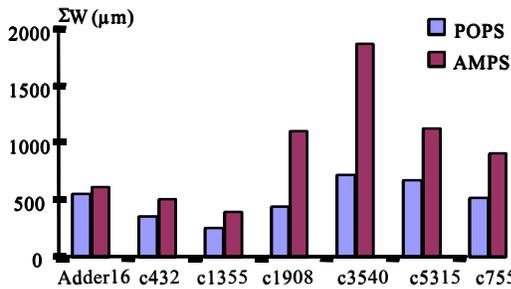

**Fig.4. Comparison of the constraint distribution methods on different ISCAS circuits.**

In Table 1 we compare the CPU time necessary for AMPS and POPS in sizing under delay constraint different benchmarks.

| Circuits | Gate nb | POPS (ms) | AMPS (ms) |
|---|---|---|---|
| Adder16 | 99 | 159 | 23700 |
| fpd | 14 | 19 | 6120 |
| c432 | 29 | 29 | 9950 |
| c499 | 29 | 30 | 9050 |
| c880 | 28 | 29 | 9850 |
| c1355 | 30 | 49 | 11400 |
| c1908 | 44 | 49 | 11760 |
| c3540 | 58 | 69 | 15890 |
| c5315 | 60 | 90 | 19400 |
| c6288 | 116 | 210 | 21920 |
| c7552 | 47 | 69 | 16400 |

**Table 1. CPU time comparison in satisfying path delay constraint.**

As illustrated the use in POPS of a deterministic approach, results in a two order of magnitude speed up of the constraint distribution step.

The second situation to be considered corresponds to delay constraint value smaller than the minimum delay available. In this case the only alternative is to modify the structure of the path.

## 4. Optimization with structure modification

The goal of this part is to set up a way to select between sizing, buffer insertion and path logic structure modification. We just focus in part 4.1 on the buffer insertion method, that will be easily extended to the logic path modification (part 4.2). The problem is to determine, at minimum area cost, the best location to insert a buffer and the minimal sizing satisfying the delay constraint.

### 4.1. Metric for path acceleration with buffer insertion

The problem here is to determine which specific node on the path must be sped up and what is the best alternative: transistor sizing or buffer insertion. In Fig 5 we represent a path general situation where an overloaded node is guessed to be sped up by buffer insertion. We intend to remove the guess by a metric directly determining for what level of load a gate switching speed can be improved. For that we compare the delay (1) of the A and B structures for determining at what fan out value ($F = C_L/C_{IN}$) the B structure becomes faster than A. This defines the "load buffer insertion limit" (Flimit). In a first step we use a local insertion method in which we conserve the size of gates (i-1) and (i) and just size the buffer (4), for minimizing the delay between the output of (i) and the terminal load.

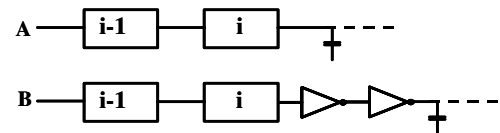

**Fig.5. Local buffer insertion.**

The values of the resulting Flimit are listed in Table 2. For the configuration given in Fig.5, (i-1) is an inverter and we have considered the evolution of the limit with the gate (i). A complete characterization must involve all possibility of (i-1) gate and can be done easily following the same procedure. Validation of these limits has been obtained through Hspice simulations. As expected, greater is the logical weight of the gate, lower is the limit that may constitute a measure of the gate efficiency. For example the Nor3 gate is the less efficient one, it must me sped up at much lower load than the other ones.

| Gate i-1 | Gate i | Calcul. | Simulation |
|---|---|---|---|
| inv | inv | 5.7 | 5.9 |
| inv | nand2 | 4.9 | 5.4 |
| inv | nand3 | 4.5 | 5.2 |
| inv | nor2 | 3.8 | 3.5 |
| inv | nor3 | 2.7 | 2.5 |

**Table 2. Fanout limit (Flimit) or a gate (i) controlled by an inverter.**

In fact the buffer insertion acts as a load dilution for the initial gate. In this case the size of this gate can be decreased. In a complete optimization flow these predefined limits are used for critical nodes identification. Then the constant sensitivity method is used to distribute the delay constraint on the full path, insuring an area efficient gate sizing.

Validation of this approach is given in Table 3 where we compare the minimum delay obtained, from POPS, on the different ISCAS circuits using sizing and buffer insertion techniques. As shown, depending on the path structure significant minimum delay value improvement can be obtained with buffer insertion. Note that considering the delay sensitivity to the gate sizing (Fig.4), any minimum delay improvement on a path will result in a delay constraint satisfaction with smaller area.

| Circuits | Method | Tmin(ns) | Circuits | Method | Tmin(ns) |
|---|---|---|---|---|---|
| Adder | sizing | 4,53 | c1908 | sizing | 2,66 |
|  | buff | 4,39 |  | buff | 2,32 |
|  | gain | 3% |  | gain | 15% |
| c432 | sizing | 2,22 | c3540 | sizing | 3,29 |
|  | buff | 1,97 |  | buff | 3,21 |
|  | gain | 13% |  | gain | 2% |
| c499 | sizing | 1,79 | c5315 | sizing | 3,57 |
|  | buff | 1,64 |  | buff | 3,20 |
|  | gain | 9% |  | gain | 12% |
| c880 | sizing | 2,09 | c6288 | sizing | 7,98 |
|  | buff | 1,71 |  | buff | 7,74 |
|  | gain | 22% |  | gain | 3% |
| c1355 | sizing | 2,16 | c7552 | sizing | 3,08 |
|  | buff | 1,89 |  | buff | 2,60 |
|  | gain | 14% |  | gain | 18% |

**Table 3. Comparison of sizing and buffer insertion techniques.**

This is illustrated in Fig.6 were we compare, on a 13 gate array, the path delay versus the area for the two methods: gate sizing (full line) and buffer insertion with global gate sizing (dotted line).

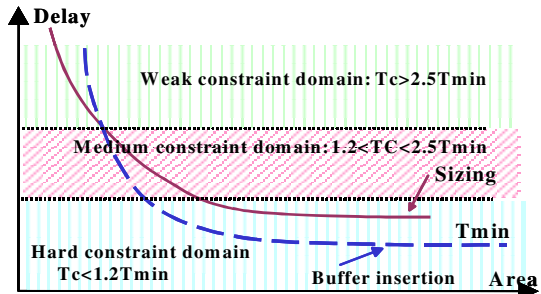

**Fig.6. Constraint domain definition.**

Three regions can be defined, a weak constraint domain where sizing is the best solution (Tc > 2.5Tmin), a medium constraint domain where buffer insertion is not necessary, but allows path implementation with area reduction (1.2Tmin < Tc <2.5Tmin) and a hard constraint domain (Tc < 1.2Tmin), where buffer insertion is the most efficient alternative.

The resulting optimization protocol, (Fig.7), has been implemented in POPS for validation on the ISCAS benchmarks. The comparison of the different steps is illustrated in Fig.8 where for three different delay constraint values (weak, medium, hard) we compare the path implementation area on the ISCAS circuits.

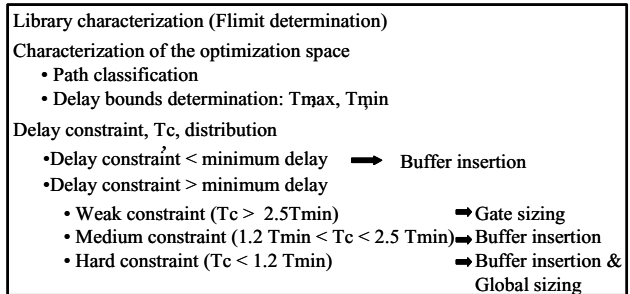

**Fig.7. Optimization protocol.**

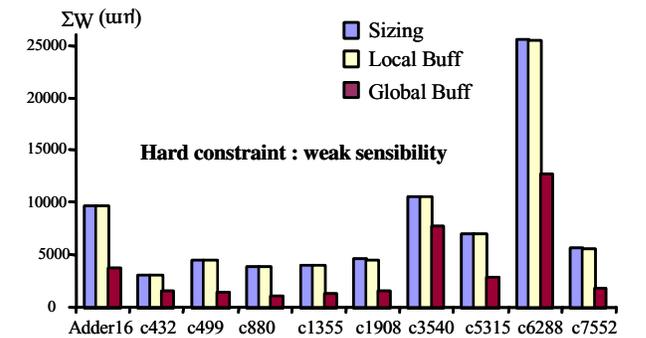

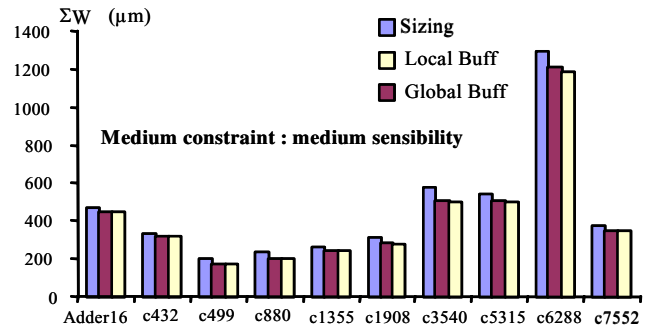

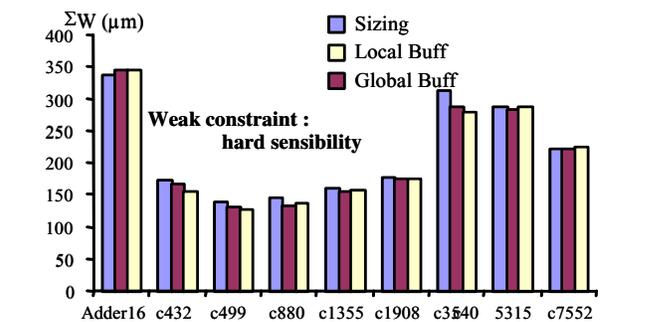

**Fig.8. Area saving in the different constraint domain for different optimization methods.**

As shown, if for weak and medium constraints the different optimization methods are quite equivalent in terms of area, for hard constraint the buffer insertion with global sizing always results in an important area saving.

### 4.2. Path acceleration with logic structure modification.

As discussed in section 4.1, the load limit for buffer insertion is a direct measure of the gate efficiency. Smaller is this limit value (Table 2) less efficient is the gate, which becomes a good candidate for buffer insertion. Another alternative is to replace an inefficient gate by a more performing one, the necessary inverters used to conserve the logic function insuring the beneficial load dilution. This corresponds to the speed up procedure currently used in path logic structure modification. Instead to speed up a gate with low sensitivity (NOR) with transistor sizing or buffer insertion we use the De Morgan's theorem to replace this gate by a more efficient one (NAND). The number of inserted inverters is the same but the second solution appears less expensive in terms of speed or area. We applied this technique to the preceding ISCAS benchmarks. In Table 4 we compare, for hard and medium timing constraints the area obtained using buffer insertion or path structure modification. As shown deterministic logic structure modification on critical path supplies a non negligible area (Power) save.

Hard constraint

| Circuits | Method | $\Sigma$ W (µm) |
|---|---|---|
| c1355 | buff | X |
| | restruct | X |
| | gain | X |
| c1908 | buff | 1522 |
| | restruct | 1286 |
| | gain | 16% |
| c5315 | buff | 2848 |
| | restruct | 2547 |
| | gain | 11% |
| c7552 | buff | 1770 |
| | restruct | 1578 |
| | gain | 11% |

Medium constraint

| Circuits | Method | $\Sigma$ W (µm) |
|---|---|---|
| c1355 | buff | 240 |
| | restruct | 230 |
| | gain | 4% |
| c1908 | buff | 280 |
| | restruct | 250 |
| | gain | 11% |
| c5315 | buff | 500 |
| | restruct | 472 |
| | gain | 6% |
| c7552 | buff | 344 |
| | restruct | 325 |
| | gain | 6% |

**Table 4. Comparison between buffer insertion and logic structure modification**

### 5. Conclusion

In circuit path optimization three major techniques are implied: transistor sizing, buffer insertion and logic transformation. Without robust indicators the selection of the more appropriate technique may lead to numerous iterations and result in transistor oversize and increase in power dissipation.

Based on a realistic model for gate timing performance, we have defined metrics for selecting path optimization alternatives. We have proposed a method for characterizing the speed possibility of a path and the gate efficiency. We have defined the minimum delay, Tmin, achievable on a path. Then we have determined, at the gate level, the fan out limit for buffer insertion, Flimit. These two indicators have been used for determining the path critical nodes (Flimit) and for selecting (Tmin) between sizing and buffer insertion alternatives.

Then based on a gate sensitivity factor "a", we have proposed a novel delay constraint distribution method, allowing path optimization at provably minimum area cost. These metrics have been used to define a general path optimization protocol that has been implemented in an optimization tool.

Validation on various benchmark circuits, implemented in a 0.25µm CMOS process, has demonstrated the validity of the defined metrics for selecting between the different optimization alternatives.